\documentclass[final,3p,times,review,authoryear]{elsarticle}

\usepackage{isotope}
\usepackage{booktabs}
\usepackage{amsmath}
\usepackage{hyperref}
\usepackage{lmodern}
\usepackage[T1]{fontenc}

\usepackage{amssymb}
\usepackage{lineno}
\usepackage{soul}
\usepackage{xcolor} % This package is used for color definitions

\sethlcolor{white} % Set the highlight color to yellow

\journal{Journal of Cleaner Production} %Journal of Cleaner Production

\begin{document}

\begin{frontmatter}

\title{Simplified Efficiency Calibration Methods for Scintillation Detectors Used in Nuclear Remediation}

\author{Victor V. Golovko \corref{cor1}} 
\ead{victor.golovko@cnl.ca}
\cortext[cor1]{Corresponding author}

\address{Canadian Nuclear Laboratories, 286 Plant Road, Chalk River, ON, Canada, K0J~1J0}

\begin{abstract}
Our study introduces innovative methods to address waste reduction and enhance resource efficiency in nuclear cleanup processes. We focus on sustainable nuclear technology, aligning with goals for cleaner production and environmental protection. We developed two novel methods, termed ``oversimplified'' and ``simplified,'' for easily determining the photopeak efficiency of NaI(Tl) scintillation detectors. These methods, along with a ``general'' solution method, were validated with calibrated radioactive sources such as $^{241}$Am, $^{57}$Co, $^{133}$Ba, $^{137}$Cs, and $^{60}$Co, and were used to commission a NaI(Tl) scintillation detector system at Chalk River Laboratories (CRL) for nuclear remediation. The system demonstrated high accuracy, making it suitable for screening unstable isotopes at contaminated sites. This screening tool significantly reduces the number of soil samples requiring detailed characterization, thereby lowering operational costs. The NaI(Tl) detector system was calibrated for near-contact geometry, which is commonly used at CRLs. Detection limits were established for this configuration. By improving the efficiency calibration of scintillation detectors, our study advances sustainable nuclear remediation practices, promoting environmental sustainability and cleaner production processes.
\end{abstract}

%Graphical abstract
\begin{graphicalabstract}
\includegraphics[width=\textwidth]{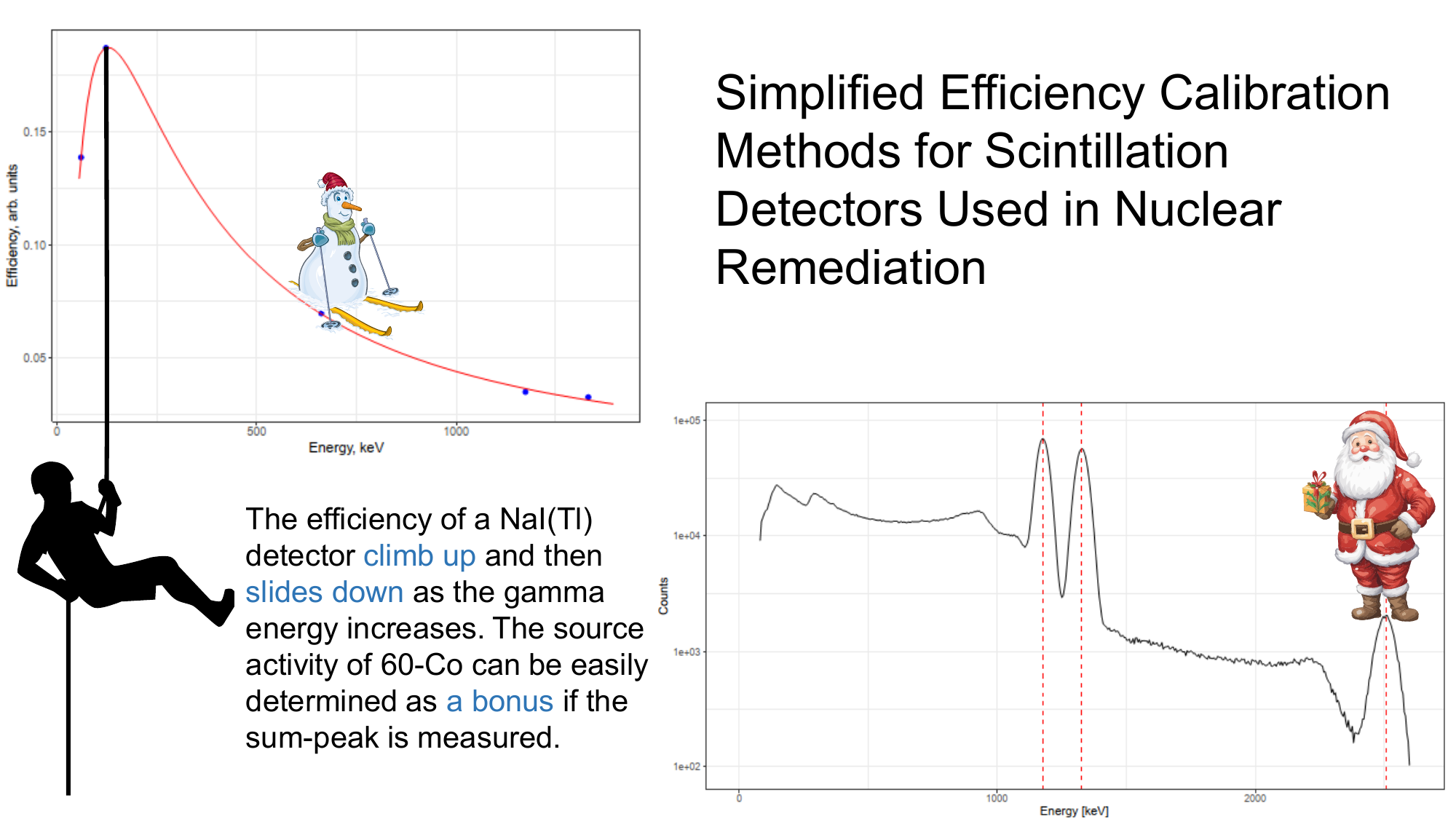}
\end{graphicalabstract}

%Research highlights
\begin{highlights}
\item The developed ``oversimplified'' and ``simplified'' calibration methods enhance the efficiency determination of NaI(Tl) scintillation detectors, contributing to more sustainable nuclear technology and cleaner production practices. These methods are validated with various radioactive sources and align with environmental sustainability goals.
\item The study demonstrated that the calibrated NaI(Tl) detector system meets all acceptance criteria for near-contact geometry, ensuring reliable screening of radioactive isotopes at contaminated sites. This reduces the amount of radioactive waste requiring detailed characterization, thereby lowering operational costs for nuclear remediation.
\item The research effectively employs R, a noncommercial and open-source data analysis tool, to analyze data from scintillation detectors. This approach not only reduces software licensing costs but also leverages the extensive verification processes of the broader scientific community, enhancing the reliability and accuracy of nuclear remediation activities.
\end{highlights}


\end{frontmatter}

%\linenumbers

%% main text
\section{Introduction}
\label{sec:intr}

Nuclear remediation is  a nuclear cleanup method that addresses issues such as radioactive contamination, especially  legacy nuclear waste. The goal is  to make contaminated areas safe by removing or reducing contamination and managing nuclear waste, thereby protecting the environment and public health.

Commissioning radiation detectors is vital for this process. It ensures safe and accurate radiation measurements, regulatory compliance, efficient cleanup, reliable data, and quality control. These detectors are crucial for effective nuclear remediation.

Thallium-activated sodium iodide detectors (NaI(Tl)) are the perfect choice for nuclear cleanup. They efficiently detect low radiation levels, cover a wide energy range, offer satisfactory energy resolution, and have various sizes. The durability of these materials  makes them suitable for challenging environments. With established calibrations, these detectors are cost-effective, maintain low background radiation levels, and have a reliable operation history.

The commissioning process verifies the performance of detectors such as NaI(Tl), ensuring that they meet standards (\cite{ANSI1994}) and comply with regulations (\cite{CNSC2021}). It also provides a baseline for future evaluations, helps with troubleshooting and maintenance, and serves as a reliable record for audits and quality management.

The acceptance criterion that has been chosen for the nuclear remediation screening tool used in this work is the requirement specifications a detector must meet to gain approval for use. For example, a NaI(Tl) detector might need to meet criteria such as an energy calibration range from 122 keV to 2,505~keV with an accuracy of 15\%, an energy resolution of 10\% at 662 keV, and an activity measurement accuracy of 15\%. 
A calibrated detector with energy up to 2.5~MeV is useful  because certain naturally occurring radioactive materials emit radiation above 2.5~MeV. For example, an intriguing gamma-ray spectrum was observed via a similar NaI(Tl) detector at specific locations while  a walking survey was conducted on one of the campuses of Hirosaki University in Japan: high-energy gamma rays from $^{208}$Tl were detected at approximately 2.6~MeV (\cite{ijerph20032657}).

The focus is on the sensitivity of NaI(Tl) detectors for rapid activity determination via field measurements. Field tools prioritize speed and broad energy detection, whereas lab detectors aim for detailed and accurate measurements in controlled settings.

In essence, a commissioning process ensures that a radioactive detector system for nuclear remediation activities operates correctly, complies with regulations, and delivers reliable measurements. It is a tool for upholding safety and quality in radiation practices.

In the field of nuclear remediation, accurately determining the photopeak efficiency of scintillation detector systems is crucial. Various methods have been developed to estimate this efficiency, each with its strengths and weaknesses. Traditional methods often involve complex calculations and extensive calibration procedures. In this paper, we introduce two new methods, termed ``oversimplified'' and ``simplified,'' designed to streamline the efficiency determination process. These methods have been validated with calibrated radioactive sources, such as $^{241}$Am, $^{57}$Co, $^{133}$Ba, $^{137}$Cs, and $^{60}$Co.

A comprehensive comparison of existing methods and our proposed methods is provided in Section~\ref{sec:Ener_Cal}. This comparison highlights the advantages of our methods in terms of accuracy, ease of implementation, and cost-effectiveness. By addressing the limitations of current approaches, our methods offer a reliable and efficient solution for nuclear remediation at Chalk River Laboratories and similar facilities.

In our study, we explore noncommercial data analysis tools such as R (\cite{rcoreteamLanguageEnvironmentStatistical2022}). 
R is a programming language that is open-source and free. It was created in 1993 by statisticians for statisticians. R is well regarded for its ability to perform statistical computations effectively (\cite{suttonStatisticsSlamDunk2024}).
Noncommercial tools save costs, offer flexibility, and are transparent. They can be customized to fit specific needs and support open science through  the sharing of methods. However, they might lack the user-friendliness and support found in commercial options. This work helps  address this gap.

\section{Methods and Materials}

To ensure precise radiation measurements during commission, specific materials and software were employed:
a cylindrical NaI(Tl) scintillation crystal (3'' diameter, 3'' height) within a Canberra detector (model NAIS-3x3),
an Osprey multichannel analyzer (tube base model Osprey-DTB), Windows 10 laptop running Genie$ \textsuperscript{\texttrademark} $~2000 (\cite{canberraGenie2000Spectroscopy2006}) Spectroscopy Software (9233652J V3.4), and
a set of standard calibrated radioactive gamma sources consisting of $^{241}$Am, $^{57}$Co, \isotope[133]{Ba} $^{137}$Cs, and $^{60}$Co.

Following the manufacturer's guidelines is essential for precise sodium iodide detector system operation. This involves instrument calibration using a radionuclide-specific standard source and verifying the absence of other radionuclides.

Maintaining the traceability of standard gamma sources is essential for energy and efficiency calibration, to establish a clear link to international standards. This ensures consistent and reliable calibration, precise radiation level measurements, and the ability to compare instruments across laboratories.

\section{Energy and Efficiency Calibration }
\label{sec:Ener_Cal}

The calibration method used for the NaI(Tl) detector during commission is based on the same principles and techniques previously utilized to calibrate the high-purity germanium detector for criticality dosimetry. Details of the method are outlined in~\cite{golovko2022simplified}.

\subsection{Near-contact Geometry}

``Near-contact geometry'' means placing a radioactive source very close to a sodium iodide detector. This setup increases gamma-ray detection efficiency, especially with weak sources, for increased accuracy. However, it can introduce challenges such as detector dead time (for strong sources) and self-absorption effects, which must be considered in data analysis.

\subsubsection{Energy Calibration for Near-contact Geometry}

For  energy calibration, the NaI(Tl) detector faces a radiation field generated by a mixed radionuclide standard that incorporates $^{241}$Am, $^{57}$Co, $^{137}$Cs, and $^{60}$Co (\cite{unknown-author-2022}). The irradiation was performed head-on, with the mixed radionuclide standard mounted on a low-scatter support, its bottom surface is directed toward the radiation source's axis. Measurements were taken at a 0.5 cm distance from the standard surface.

\begin{figure*}[t]
	\centering
	\includegraphics[width=\textwidth]{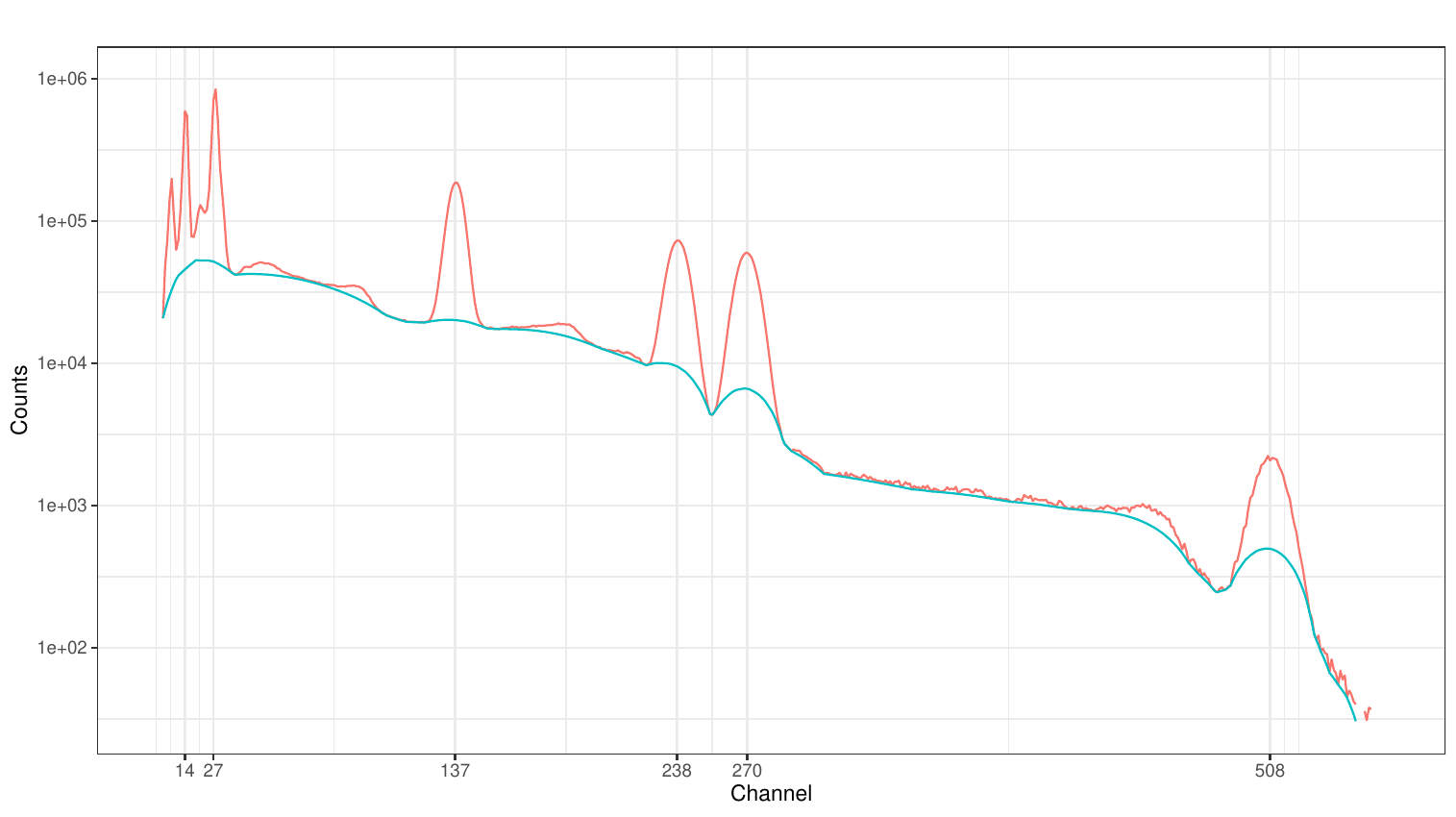} \\[-10pt]
	\caption{A raw spectrum of a multinuclide calibration source, including $^{241}$Am, $^{57}$Co, $^{137}$Cs, and $^{60}$Co, was obtained via a NaI(Tl) detector. The measurement was performed with a live time of 10 minutes. The spectrum covers the energy range up to the true coincidence sum peak of $^{60}$Co at channel 508. }
	\label{fig:NaI_mult_ch}
\end{figure*}

Before measurement, the multichannel analyzer and detector were powered on and allowed to stabilize for a minimum of 15 minutes. Each measurement was repeated 50 times, each lasting 10~minutes. The raw spectrum from a 10-minute measurement of a multinuclide calibration source is displayed in Figure~\ref{fig:NaI_mult_ch}. This spectrum depicts the energy distribution of detected gamma rays, showing 560 of the 1,024 channels.

The subsequent stage involved analyzing the spectra to identify the energy peaks corresponding to the emitted gamma rays. Specialized software or tools such as R-Soft were employed for peak identification. We utilized the {\tt{gamma}} package from R (\cite{lebrunGammaPackageDose2020}), which is designed to analyze gamma-ray spectrometry measurements and is compatible with raw Canberra data files. The package's functionalities and techniques for managing gamma-ray spectra are detailed in other sources (\cite{ryan1988snip, morhavc1997background, morhavc2008peak}). This package provides various tools for processing and analyzing gamma-ray data efficiently.

In Figure~\ref{fig:NaI_mult_ch}, baseline subtraction was utilized exclusively for peak identification, not for precise count determination. Alternatively, a manual approach visually inspects channels with the highest counts to identify energy peaks. After  the peaks were identified  (refer to Figure~\ref{fig:NaI_mult_ch} and Table~\ref{tab:mu_su}), the next step was to establish a calibration curve or equation.

The calibration curve relates channel numbers or voltage levels recorded by the detector to corresponding gamma-ray energies. It accurately determines gamma-ray energy  on the basis of the channel number. The curve is obtained by fitting a mathematical function to previously identified energy peaks, facilitating reliable energy determination in subsequent measurements.

\begin{table*}[t]
	\centering
	\caption{Summary of region of interest (ROI) peaks shown for a mixed radionuclide standard, which included $^{241}$Am, $^{57}$Co, $^{137}$Cs, and $^{60}$Co. The measurement took place on April 6, 2023, and started at 2:34:58 PM. The test was conducted for a total live time of 10 minutes. }	
	\begin{tabular}{cccccccccc}
		\hline
		\\ [-10pt]
		ROI & Nuclide & $ E $ & Centroid & Ch & ROI  & $ N_m(E) $ & $ \sigma_i(E) $ & FWHM \\
		&  & keV & keV  &   & Counts & Counts  & Counts & keV \\
		\hline
		\\ [-10pt]
		1 & $^{241}$Am & 59.54 & 57.03 & 14 & 1,914,567 & 1,305,782 & 1,355  & 10.13 \\
		2 & $^{57}$Co  & 122.06 & 121.80 & 27 & 3,832,566 & 2,347,068 & 1,765  & 14.80 \\
		3 & $^{137}$Cs & 661.66 & 669.16 & 137 & 2,058,850 & 1,537,147 & 1,525  & 42.42 \\
		4 & $^{60}$Co  & 1,173.23 & 1,170.71 & 238 & 1,023,254 & 794,738 & 1,116  & 55.14 \\
		5 & $^{60}$Co  & 1,332.49 & 1,329.40 & 270 & 842,039 & 743188 & 1,046  & 60.53 \\
		6 & $^{60}$Co  & 2,505.69 & 2,506.57 & 508 & 46,182 & 35,543 & 219   & 81.18 \\
		\hline
	\end{tabular}%	
	\label{tab:mu_su}%
\end{table*}%

Energy calibration involves fitting a channel number (Ch) to gamma-ray energy ($E$) via a quadratic equation that accordingly includes the coefficients for the intercept, linear, and quadratic terms:
\begin{equation}\label{eq:En_Ch}
	E = a_0 + a_1 \cdot \text{Ch} + a_2 \cdot \text{Ch}^2.
\end{equation}

\begin{figure*}[t]
	\centering
	\includegraphics[
	width=\textwidth]{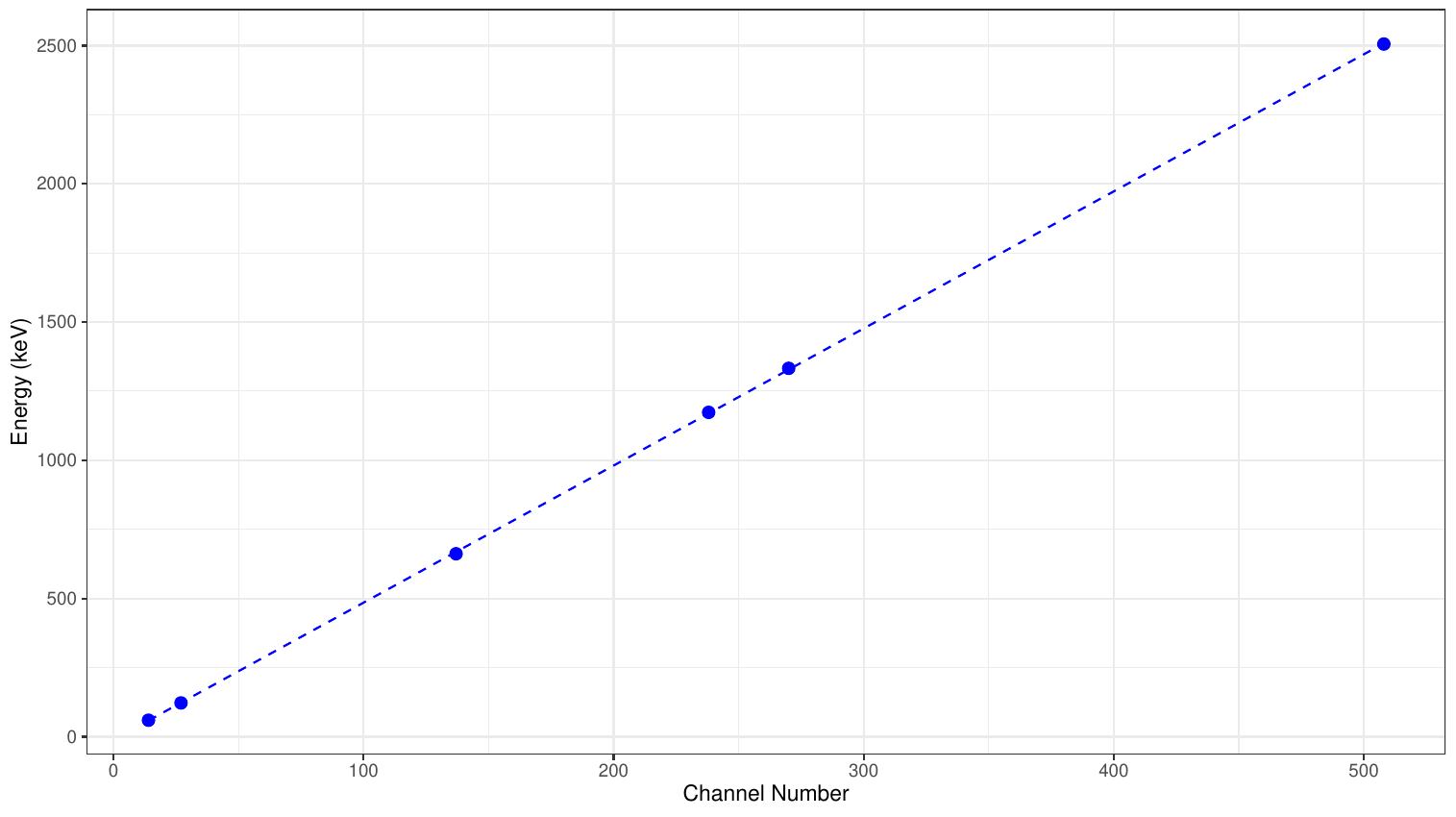} \\ [-10pt]
	\caption{Energy calibration curve created via gamma rays from a mixture of radioactive nuclei such as $^{241}$Am, $^{57}$Co, $^{137}$Cs, and $^{60}$Co. This curve is for a NaI(Tl) detector and covers energies from approximately 60 keV up to channel 508, where we find the true coincidence sum peak of $^{60}$Co. }
	\label{fig:en_ch_cal}
\end{figure*}

Although more complex fits are possible, they are not theoretically justified, as these systems are designed for linearity. Therefore, a quadratic fit should be sufficient for calibrating a thallium-activated sodium iodide detector. An analog-to-digital converter converts analog signals to digital data.
The specific fit for Equation~\ref{eq:En_Ch} is  as follows:
\begin{equation}\label{eq:en_ch_fit}
	E = -12.7 + 4.98 \cdot \text{Ch} - 0.00005 \cdot \text{Ch}^2.
\end{equation} 
As shown  in Figure~\ref{fig:en_ch_cal}, this equation illustrates the energy-to-channel relationship.

The energy calibration range, from 60 keV to 2.5~MeV, is crucial for accurate measurements. The fitted values for each nuclide's energy centroid (central values) are compared with the actual values to assess the calibration  accuracy. For example, the predicted energy for a 59.54 keV gamma ray from $^{241}$Am is 4\% lower than its true value (see Table~\ref{tab:mu_su}), meaning that  measurements should fall within 4\% of the expected range. This 4\% range offers confidence in the measurement accuracy. Similarly, for a 2,505.69 keV gamma ray from $^{60}$Co, the predicted energy is less than 1\% (see Table~\ref{tab:mu_su}). This demonstrates that the calibration meets the ``within 15\% accuracy'' criterion  over this energy range.

\begin{figure*}[t]
	\centering
	\includegraphics[width=\textwidth]
	{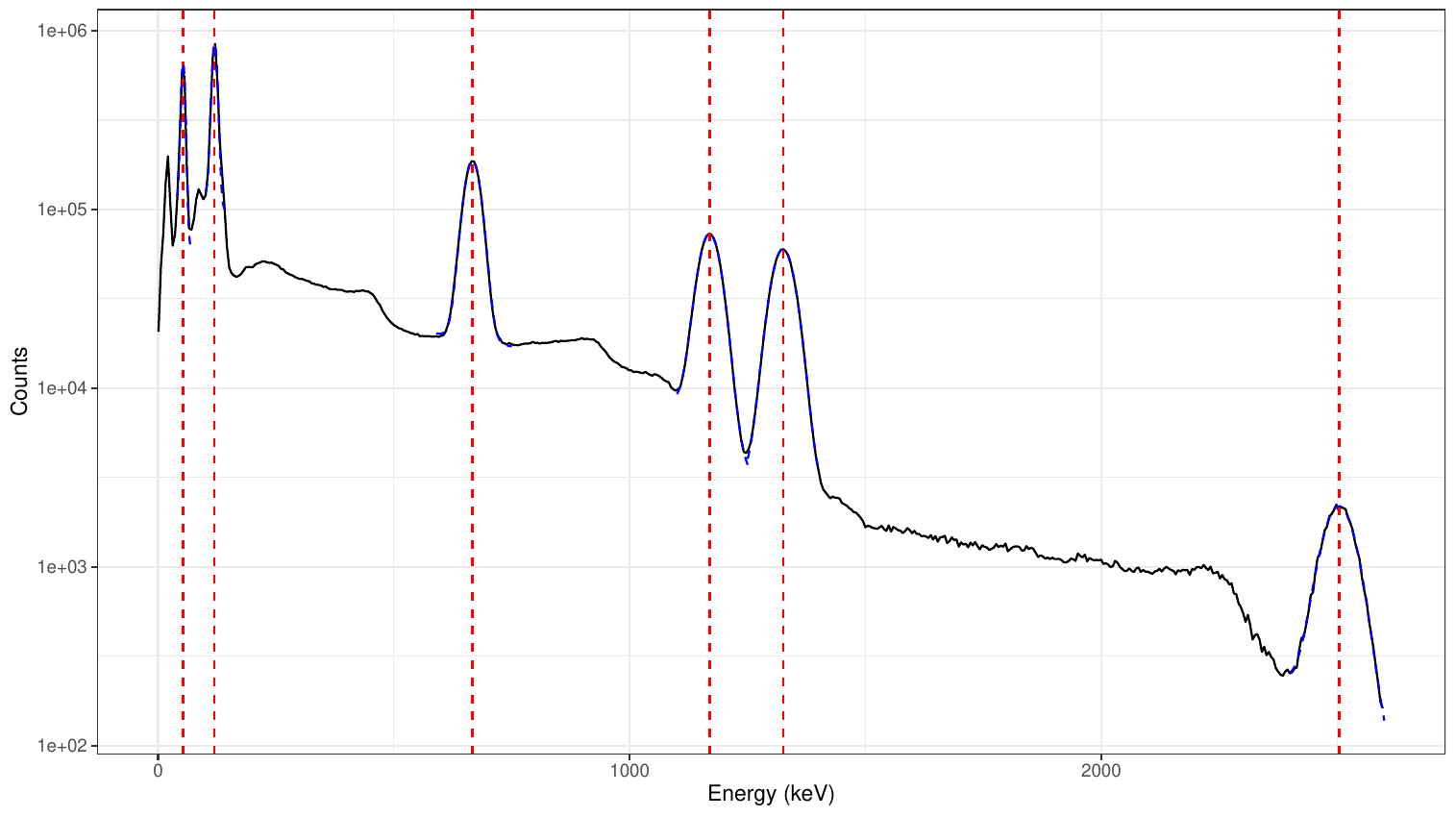} \\[-10pt]
	\caption{A mixture  of radioactive nuclei, such as $^{241}$Am, $^{57}$Co, $^{137}$Cs, and $^{60}$Co, was used to generate  a calibration spectrum. We used a NaI(Tl) detector and measured it for 10 minutes. This spectrum shows energies up to 2.5 MeV, which is the true coincidence sum peak for $^{60}$Co. }
	\label{fig:NaI_mult}
\end{figure*}

Figure~\ref{fig:NaI_mult} shows the graph of the multinuclide calibration source after calibration. The figure  shows the distinct photopeak lines. We calculated the net count area by removing the background and fitting the data with a Gaussian function. This area is reported in Table~\ref{tab:mu_su}, along with its uncertainty.

For the 662 keV photopeak line from $^{137}$Cs, the full width at half maximum (FWHM) was 42.42~keV, which translates to an accuracy of 6.34\%. This accuracy is better than the required 10\% criterion for  $^{137}$Cs photopeak resolution.

\subsubsection{Efficiency Calibration for Near-contact Geometry}

We can calibrate the efficiency of the NaI(Tl) detector system via the cascade summing phenomenon. \cite{golovko2022simplified} proposed simplified methods to take advantage of the true coincidence summing effect for semiconductor detectors. These methods are also suitable for scintillation detectors such as NaI(Tl). By using these methods with a calibrated $^{60}$Co source, we can commission any NaI(Tl) detector system at Chalk River Laboratories (CRL). The accuracy achieved is reasonable and sufficient for nuclear remediation detector commissioning within acceptance criteria. 

The handbook of radioactivity measurement procedures (\cite{national1985handbook}) explains the sum--peak coincidence counting method used to measure activity during radioactive decay. This method is also used for the absolute standardization of radioisotopes by employing a single-crystal NaI(Tl) detector system, achieves  an accuracy of approximately $\pm$5\%, as mentioned by \cite{brinkman1963absolute}. In our study, we  also focused on considering the angular correlation between the two gamma rays in the decay of the $^{60}$Co radioactive source to describe the activity of the radioisotope. Furthermore, we  investigated the ability to measure both the activity of $^{60}$Co and the photopeak efficiency of the detector in the presence of other radioisotopes. This approach is also relevant in the calibration of semiconductor detectors (\cite{golovko2022simplified}).

An improved formula was developed by \cite{nemes2016improved} for determining the activity of a two-photon cascade emitting source via the sum--peak method. According to this method, the weighted mean deviation of the measured activity from the reference value of a $^{60}$Co source was found to be approximately 0.1\%. Later, \cite{nemes2018generalized} generalized this method to account for random coincidences. 

A review of sum--peak methods for measuring absolute source activity via a single detector, along with corrections for random summations, was provided by \cite{MALONDA2020531}. However, there are typographical errors in equations 7.64 and 7.65. These errors were noted in the book edited by \cite{lannunziataRadioanalyticalApplicationsVolume2020}. Interestingly, the previous edition of the book, edited by \cite{lannunziataHandbookRadioactivityAnalysis2012}, does not contain these typos in the theoretical description of the sum--peak method reviewed by \cite{MALONDA2012871}.

Importantly, to measure the activity of $^{60}$Co with a precision of 0.1\% (\cite{nemes2016improved}), the dead time of the detector was more than 35\%, and in certain measurements, it even reached 75\%. Such a high percentage of dead time is unexpected for nuclear remediation work, particularly when working with small amounts of radioactive isotopes. In this study, we used a method based on the true coincidence sum--peak approach originally developed for criticality dosimetry with semiconductor detectors (\cite{golovko2022simplified}) and applied it to scintillation detectors.

\subsubsection{``Oversimplified'' Solution for Near-contact Geometry} \label{sec:ovs}

Using the same notations as in \cite{golovko2022simplified}, we estimate the ``\text{oversimplified}'' solution for the activity of a calibrated $^{60}$Co radioactive source and the efficiencies for the corresponding gamma lines via the following set in Equation~\ref{eq:ovs_sol} (see Equation~\ref{eq:sp_thos} in~\ref{ap:theory}):
\begin{equation}
	\label{eq:ovs_sol}
	\begin{split}
		A & = \frac { N_1 \, N_2 \, \omega(0^{\circ}) } { N_{12} }  \\
		\varepsilon_1 & = \frac {N_{12}}{N_2 \, p_1 \, \omega(0^{\circ})} \\
		\varepsilon_2 & = \frac {N_{12}}{N_1 \, p_2 \, \omega(0^{\circ})}. 
	\end{split}
\end{equation}
Here, the full-energy peak areas are $N_1$,  $N_2$, and $N_{12}$; $p_1$ and $p_2$ are the photon emission intensities; and $\omega(0^{\circ})$ is the factor used  to account for angular correlations between the gamma ray photons. These equations represent the observed decays from the source during a specific time period ($A$) and the full-energy peak efficiencies of the NaI(Tl) detector system for the corresponding gamma rays in $^{60}$Co ($\varepsilon_1$ and $\varepsilon_2$). After substituting real data values from Table~\ref{tab:mu_su}, $p_1=0.9998$,  $p_2=0.9985$ (\cite{TabRad_v3}), and $\omega(0^{\circ})=1.1667$ (\cite{lemmer1954angular, golovko2022simplified}), into these equations and applying standard error analysis propagation rules (\cite{gilmorePracticalGammaraySpectroscopy2008,hughesMeasurementsTheirUncertainties2010,knollRadiationDetectionMeasurement2010}), we obtain the following results (Equation~\ref{eq:SysSolDatEr}):
\begin{equation}
	\label{eq:SysSolDatEr}
	\begin{split}
		A &= \text{19,387,766}(\text{125,535}) \\%& 
		\varepsilon_1 (\text{1,173} \, \rm{keV})  &= 0.04100(26) \\%&
		\varepsilon_2 (\text{1,332} \, \rm {kev})  &= 0.03839(24).
	\end{split}
\end{equation}
The calculated activity of the $^{60}$Co source during the observation period ($T_L = 600$ s) is $S$ (Equation~\ref{eq:S_err}): %= 32,313(209) \, \text{Bq} = 32.313(209) \, \text{kBq}$ (Equation~\ref{eq:S_err}). 
\begin{equation}
	S = \text{32,313}(209) \, \rm{Bq} = 32.313(209) \, \rm{kBq}.
	\label{eq:S_err}
\end{equation}
The activity of the $^{60}$Co source is also estimated via data from the calibration certificate~(\cite{unknown-author-2022}) and the radioactive decay law.
By combining this information, we obtain the following result (Equation~\ref{eq:dec_data_er}):
\begin{equation}
	A(t) = 37,641 (1129)\, \rm{Bq} = 37.641(1.129) \, \rm{kBq}.
	\label{eq:dec_data_er}
\end{equation}

Comparing the results from Equations~\ref{eq:S_err} and \ref{eq:dec_data_er}, we observe that the two values are within 14\% of each other. This discrepancy is attributed to the systematic uncertainties associated with our ``\text{oversimplified}'' method, as outlined in~\cite{golovko2022simplified}. Notably, this 14\% systematic uncertainty is distinct from the 1\% uncertainty arising from counting statistics, as indicated in Equations~\ref{eq:SysSolDatEr} and \ref{eq:S_err}. Despite these complexities, our activity determination falls within the specified acceptance criterion of a 15\% range.

\subsubsection{``Simplified'' Solution for Near-contact Geometry}
\label{sec:s}

Using the symbols and terminology from~\cite{golovko2022simplified}, we calculate the ``\text{simplified}'' solution for the activity of a calibrated $^{60}$Co radioactive source and the efficiencies for the associated gamma lines via Equation~\ref{eq:s_sol} (see Equation~\ref{eq:sp_ths} in~\ref{ap:theory}):
\begin{equation}
	\label{eq:s_sol}
	\begin{split}
		A & = \frac { (N_1 + N_{12}) \, ( N_2 + N_{12}) \, \omega(0^{\circ}) } { N_{12} }  \\
		\varepsilon_1 & = \frac {N_{12}}{(N_2 + N_{12}) \, p_1 \, \omega(0^{\circ})} \\
		\varepsilon_2 & = \frac {N_{12}}{(N_1 + N_{12}) \, p_2 \, \omega(0^{\circ})}. 
	\end{split}
\end{equation}

By comparing the source activity of \isotope[60]{Co} obtained via the ``\text{simplified}'' method to the source activity mentioned in the calibration certificate for the standard source, one can evaluate the accuracy of the ``\text{simplified}'' method for commissioning the NaI(Tl) detector system. The assumptions of this method are expected to yield a more accurate solution than  those of the ``\text{oversimplified}'' method.

By substituting the real data values from Table~\ref{tab:mu_su}, $p_1=0.9998$, $p_2=0.9985$, and $\omega(0^{\circ})=1.1667$, into Equation~\ref{eq:s_sol} and following standard error analysis propagation rules, the following results can be obtained:
\begin{equation}
	\label{eq:simSolDatEr}
	\begin{split}
		A &= \text{21,223,533}(\text{125,535}) \\ 
		\varepsilon_1 (\text{1,173} \, \rm{keV})  &= 0.03913(24) \\
		\varepsilon_2 (\text{1,332} \, \rm {kev})  &= 0.03675(22) \\
		S &= 35.373(210) \, \rm{kBq}.
	\end{split}
\end{equation}

A comparison with the Equation~\ref{eq:dec_data_er} result reveals  a 6\% difference,  which is  well within the 15\% acceptance criterion. A systematic 6\% uncertainty is attributed to assumptions in the ``\text{simplified}'' method (see \ref{ap:theory}) and is distinct from the 1\% counting statistics error. 
Notably, much like the outcomes observed with the semiconductor detector (\cite{golovko2022simplified}), the ``\text{simplified}'' method yields superior results for the scintillating NaI(Tl) detector  compared with the ``\text{oversimplified}'' approach.

\subsubsection{``General'' Solution for Near-contact Geometry}

We used the ``\text{simplified}'' and ``\text{oversimplified}'' methods to determine $^{60}$Co activity without prior knowledge, which demonstrated accuracy within a 15\% range. Typically, NaI(Tl) detector efficiency calibration involves selecting known gamma-ray sources, performing a stable detector setup, acquiring spectral data, identifying peaks, calculating relative efficiencies, fitting to a function, and verifying calibration. This ensures accurate efficiency determination for various gamma-ray energies, instilling confidence in the detector's performance for radiation detection and measurement. Both approaches contribute to robust activity determination, balancing simplicity with accuracy and adhering to calibration standards.

In this work, we determine the efficiency of a NaI(Tl) detector by analyzing full-energy photo peaks from radioactive gamma sources such as \isotope[241]{Am}, \isotope[57]{Co}, \isotope[137]{Cs}, and \isotope[60]{Co} (see Table~\ref{tab:mu_su}). The efficiency, denoted as $\varepsilon_i(E)$, represents the detector's ability to detect gamma rays of specific energies. This value  is calculated via the following equation:
\begin{equation}
	\varepsilon_i(E) = \frac{N_m(E) \cdot K_m}{S \ \cdot t_m \cdot p_i }.
	\label{eq:eff}
\end{equation}
Here, $N_m(E)$ is the net peak count, $t_m$ is the measurement duration, $p_i$ is the gamma-ray emission probability, and $S$ is the radionuclide activity during measurement. Typically, a correction factor ($K_m$) adjusts for decay, random summing, and true coincidence summing. However, owing  to the short measurement times and accurate corrections from our ``\text{simplified}'' method, $K_m$ is considered to be unity. The efficiencies $\varepsilon_1$ and $\varepsilon_2$ from the ``\text{simplified}'' method (or ``\text{oversimplified}''), as shown in Equation~\ref{eq:simSolDatEr} (or Equation~\ref{eq:SysSolDatEr}), can also be expressed as $\varepsilon_i(E) = \varepsilon_i(E(\text{1,172})) = \varepsilon_1$ for the 1,172 keV gamma line and $\varepsilon_i(E) = \varepsilon_i(E(\text{1,333})) = \varepsilon_2$ for the 1,333~keV gamma line. These efficiencies have already accounted for the corrections related to true coincidence summing.

The efficiency equation for gamma rays can take different forms depending on the specific model or approach used. One commonly used equation is the exponential attenuation form, represented by Equation~\ref{eq:exp_atten}:
\begin{equation}
	\varepsilon(E) = C_1 \cdot e^{-C_2 \cdot E}.
	\label{eq:exp_atten}
\end{equation}

In Equation \ref{eq:exp_atten}, $\varepsilon(E)$ represents the efficiency of the detector at a specific gamma-ray energy $E$. The parameters $C_1$ and $C_2$ are constants that are determined through calibration measurements or Monte Carlo simulations and are specific to the detector and experimental setup.
The exponential attenuation form captures the decreasing trend of detector efficiency as the energy of the gamma rays increases. 

Notably, Equation \ref{eq:exp_atten} is just one example of an explicit form of the efficiency equation, and other functional forms can be used depending on the specific characteristics of the detector and the phenomena being considered. Additionally, different types of detectors and experimental setups may require different equations or correction factors to accurately describe the efficiency.

\begin{table*}[t]
	\centering
	\caption{Efficiency of the full-energy peak, or photopeak efficiency, calculated for radionuclide mixture sources containing \isotope[241]{Am}, \isotope[57]{Co}, \isotope[137]{Cs}, and \isotope[60]{Co}, and  Equation~\ref{eq:eff} was used. References are provided for the number of characteristic photons per 100 disintegrations in these radioactive gamma sources. The radionuclide activity ($S$~Cert.) at the time of measurement is provided, along with an estimation of the efficiency error. This estimation takes into account the statistical error in the photopeak counts and a relative uncertainty of 1.5\% (or 1$\sigma$), which is roughly equivalent to a confidence level of 68\% for  radionuclide activity. }
	\begin{tabular}{cccccccrr}
		\hline
		\\ [-10pt]
		Nuclide &  $ E $ & $ N_m(E) $ & $ \sigma_i(E) $ & $S$ Cert. & $ p_i $  & References & \multicolumn{1}{c}{$\varepsilon_i(E)$} & $ \sigma_{\varepsilon_i(E)} $  \\
		& keV & Counts & Counts & Bq & $ \times 100 $ &  &  &  \\
		\hline
		\\ [-10pt]
		$^{241}$Am &    59.54 & 1,305,782 & 1,355 & 43,658 & 35.92 & \cite{TabRad_v5} & 0.1388 & 0.0021 \\ 
		$^{57}$Co &   122.06 & 2,347,068 & 1,765 & 24,437 & 85.60 & \cite{BHAT1998415} & 0.1870 & 0.0028 \\ 
		$^{137}$Cs &   661.66 & 1,537,147 & 1,525 & 43,105 & 85.10 & \cite{Browne2007} & 0.0698 & 0.0010 \\ 
		$^{60}$Co & 1,173.23 & 794,738 & 1,116 & 37,641 & 99.85 & \cite{TabRad_v3} & 0.0352 & 0.0005 \\ 
		$^{60}$Co & 1,332.49 & 743,188 & 1,046 & 37,641 & 99.98 & \cite{TabRad_v3} & 0.0329 & 0.0005 \\
		\hline
	\end{tabular}%
	\label{tab:eff_gen}%
\end{table*}%

Another form of the efficiency equation, which takes into account additional factors, is given by:
\begin{equation}
	\varepsilon(E) = e^{(A + B \cdot \ln(E) + C \cdot \ln(E)^2 + D \cdot \ln(E)^3)}.
	\label{eq:exp_additional}
\end{equation}
In Equation \ref{eq:exp_additional}, $\varepsilon(E)$ still represents the efficiency of the detector at a specific gamma-ray energy $E$. Parameters $A$, $B$, $C$, and $D$ are constants that determine the shape and characteristics of the efficiency curve, taking into account the logarithmic dependence on the energy of the gamma rays.

The specific fit for Equation~\ref{eq:exp_additional} via the ``\text{general}'' solution and data from Table~\ref{tab:eff_gen} is as follows:
\begin{equation}\label{eq:exp_coef}
	%\begin{split}
		\varepsilon(E) = e^{(-19.6 + 8.81 \cdot  \ln(E) \\
		- 1.35 \cdot \ln(E)^2 + 0.061 \cdot \ln(E)^3)}.
	%\end{split}
\end{equation} 
To obtain a slightly better result, the efficiency values for $^{60}$Co gamma lines were used from the ``\text{simplified}'' solution (see Equation~\ref{eq:simSolDatEr}) in the fit for Equation~\ref{eq:exp_additional}, while keeping the rest of the data from Table~\ref{tab:eff_gen}. The resulting equation is as follows:
\begin{equation}\label{eq:exp_coef_s}
	%\begin{split}
		\varepsilon(E) = e^{(-22.6 + 10.6 \cdot  \ln(E) \\
		- 1.72 \cdot \ln(E)^2 + 0.084 \cdot \ln(E)^3)}.
	%\end{split}
\end{equation} 

\begin{figure*}[t]
	\centering
	\includegraphics[
	width=\textwidth]{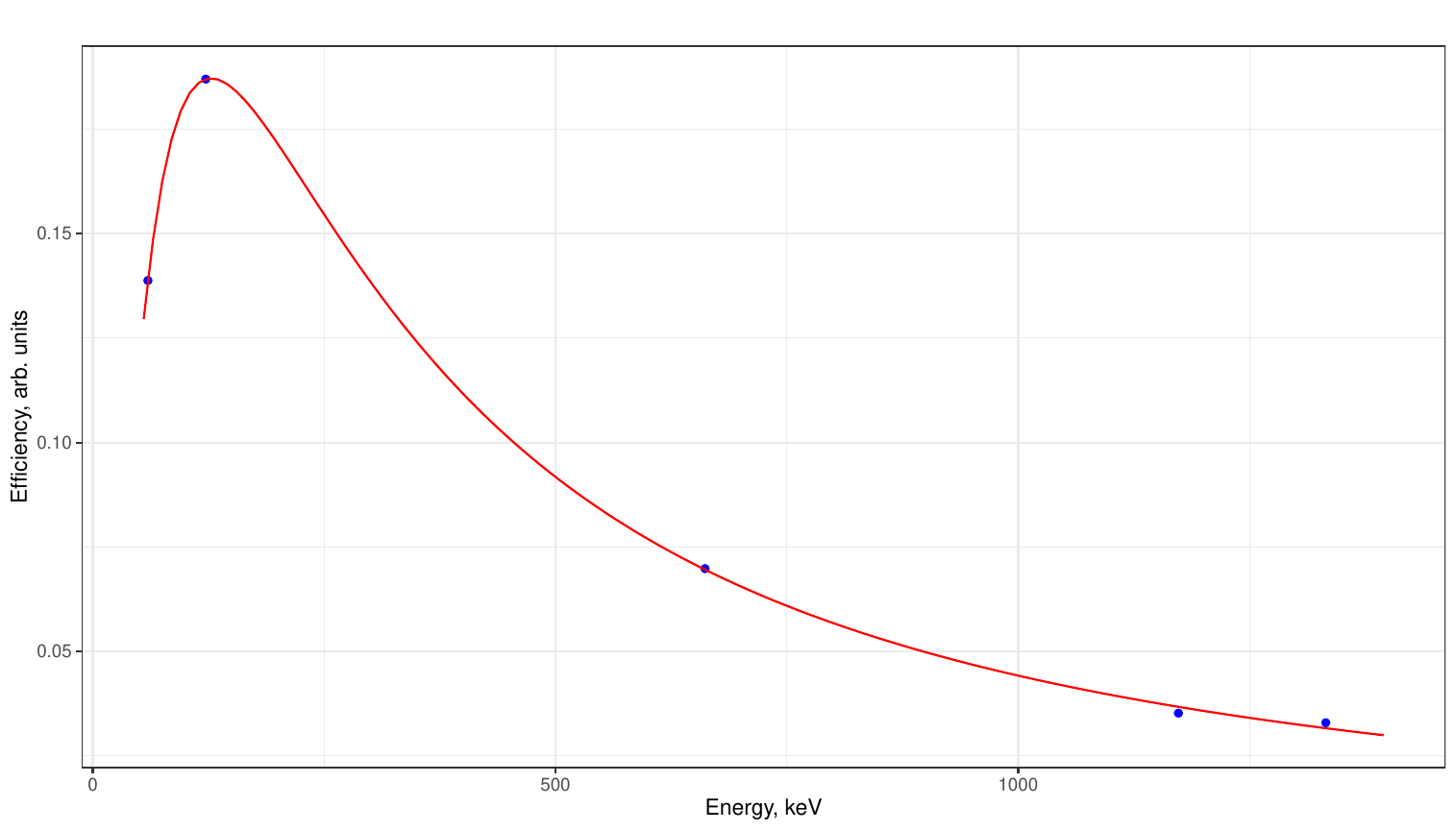} \\[-10pt]
	\caption{Efficiency curve created by fitting a combination of radionuclide standards (such as $^{241}$Am, $^{57}$Co, $^{137}$Cs, and $^{60}$Co) via Equation~\ref{eq:exp_additional} and the data presented in Table~\ref{tab:eff_gen}. This efficiency curve was established via a NaI(Tl) detector.}
	\label{fig:effNaI_en}
\end{figure*}

The efficiencies obtained via Equation~\ref{eq:eff}, which provides a ``\text{general}'' solution for each gamma-ray energy, are used as input parameters for fitting an efficiency curve (as presented in Table~\ref{tab:eff_gen}). Importantly,  the efficiency values for the characteristic gamma rays in the decay of the \isotope[60]{Co} nuclide were also calculated via Equation~\ref{eq:eff}. For more accurate results, efficiencies that account for the true coincidence summing effect for the same energies (as given by Equation~\ref{eq:simSolDatEr} in the ``\text{simplified}'' solution) can be used. The overall shape of the curve is described by Equation~\ref{eq:exp_additional}. By fitting this curve to the measured efficiencies, we can precisely estimate the efficiencies for gamma-ray energies that were not part of the calibration process. 
The outcomes of this fitting procedure are illustrated in Figure~\ref{fig:effNaI_en}.  

The precision of the efficiency for activity determination, ranging from 122 keV (for \isotope[57]{Co}) to 1,333 keV (for \isotope[60]{Co}), was further examined in this work. Notably,  the accuracy must meet the pass criterion, which requires it to be within 15\% as specified in the commissioning requirements. Thus far, we have observed that by utilizing the ``\text{simplified}'' solution for determining the activity of \isotope[60]{Co}, the accuracy of the obtained result is within 6\%, which fully meets the acceptance criterion.

\subsubsection{Validation of Activity Estimation for Near-contact Geometry}

Validating estimated activities with standard radioactive sources and a NaI(Tl) detector involves crucial steps. First, comparing estimated and measured activities ensures result accuracy and builds confidence in the estimation method. Second, it verifies the NaI(Tl) detector's calibration by comparing the measured and known activities of standard sources. Third, it acts as a quality control method, identifying systematic errors for improved accuracy. This process ensures compliance with regulatory standards and protocols, enhances  confidence in reported results for stakeholders and maintains accuracy.

From a practical standpoint,  activity validation can be carried out by calculating the activity via Equation~\ref{eq:eff} and employing the fitted efficiency curve (described by Equation~\ref{eq:exp_additional}) derived from a multisource standard. In this equation, the activity term (S) is isolated on one side of the equation, and the other terms remain the same as those in Equation~\ref{eq:eff}:
\begin{equation}
	S = \frac{N_m(E) \cdot K_m}{\varepsilon_i(E) \cdot t_m \cdot p_i}.
	\label{eq:activity}
\end{equation}
By performing measurements with another standard radioactive source of known activity under the same setup, a comparison between calculated and known activities provides an independent assessment of estimation accuracy. This serves as a validation for the methods and approximations used in determining activity results.

Additionally, the verification process can include a nuclide not used in the calibration, such as the standard source  \isotope[133]{Ba}~(\cite{unknown-author-2022}). With characteristic gamma-ray energies lying between those used in the calibration, measuring and comparing the activity to the expected activity  on the basis of the  calibration curve further validates the calibration accuracy and reliability.

On the other hand, the decay process of \isotope[133]{Ba} is more complex than  that of \isotope[137]{Cs} or \isotope[60]{Co}. Therefore, a correction factor may be necessary. However, one could also take the opposite view and ask whether it is possible to estimate the source activity via Equation~\ref{eq:activity} without considering the correction factor while still meeting the acceptance criterion for activity estimation. 

\begin{table*}[t]
	\centering
	\caption{Summary of the peaks observed in the ROI for a standard \isotope[133]{Ba} radionuclide~(\cite{unknown-author-2022}). The measurement was conducted on May 8, 2023, starting at 9:27:17 AM, and lasted for a total live time of 10 minutes. The activity of \isotope[133]{Ba} was estimated via Equation~\ref{eq:activity} and fitted efficiencies obtained from a fit of a mixture of radionuclide standards~(\cite{unknown-author-2022}) that included $^{241}$Am, $^{57}$Co, $^{137}$Cs, and $^{60}$Co. }
	\begin{tabular}{ccccccccc}
		\hline
		\\ [-10pt]
		Nuclide & $ N_m(E) $ & $ \sigma_i(E) $ & $ E $ & Centroid & Activity & $\varepsilon_i(E)$ & $ p_i $ & References \\
		& Counts & Counts & keV & keV & Bq &  & $\times 100$ & \\
		\hline
		\\ [-10pt]
		\isotope[133]{Ba} &   560,175 & 1,174 & 302.9 & 291.0 & 36,900 & 0.1380 & 18.31 & \cite{KHAZOV2011855} \\ 
		\isotope[133]{Ba} & 1,578,425 & 1,461 & 356.0 & 352.4 & 34,700 & 0.1221 & 62.05 & \cite{KHAZOV2011855} \\
		\hline
	\end{tabular}%
	\label{tab:Ba133}%
\end{table*}%

\begin{figure*}[t]
	\centering
	\includegraphics[width=\textwidth]{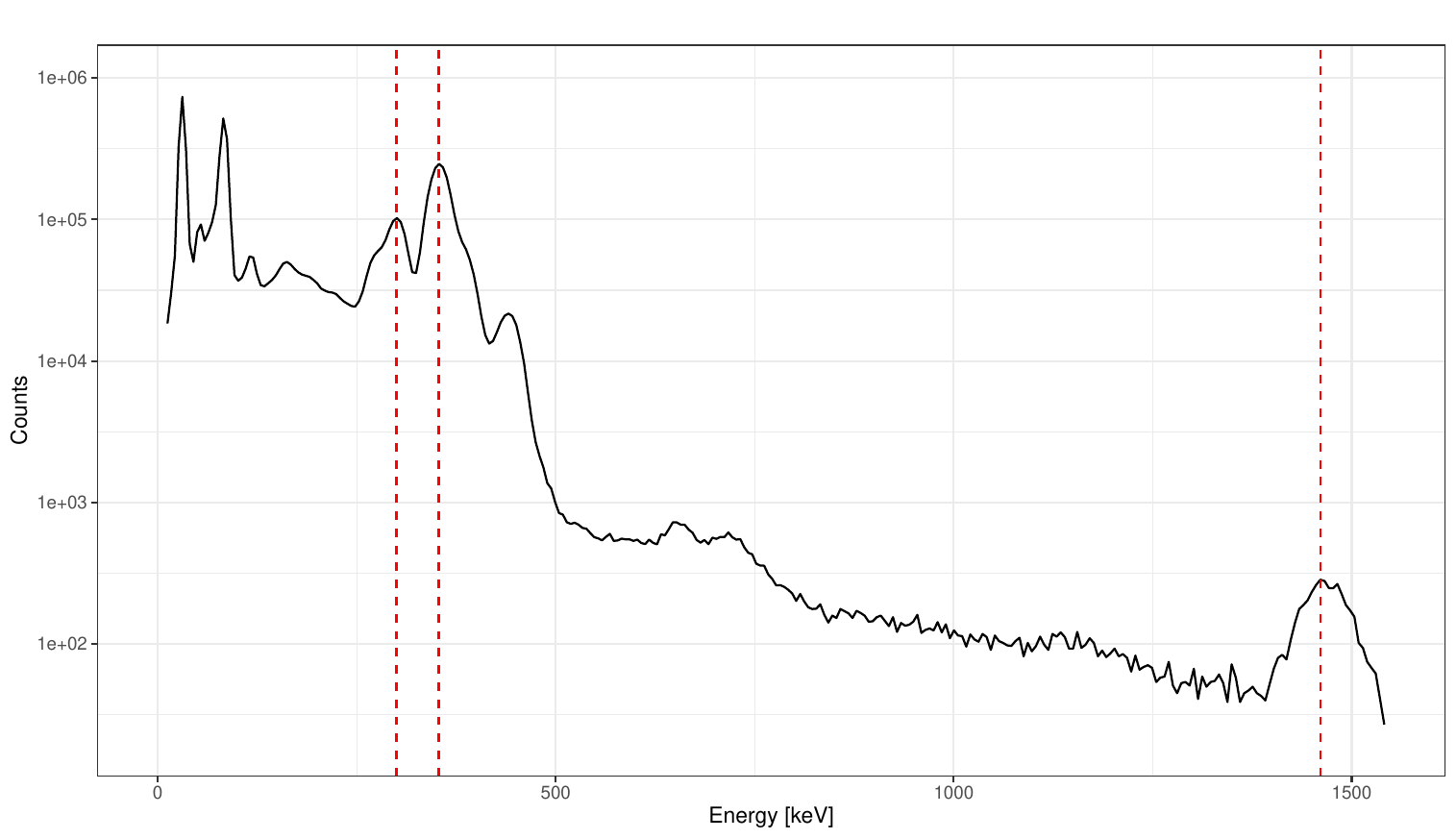} \\[-10pt]
	\caption{The spectrum of the \isotope[133]{Ba} radionuclide standard    obtained via  a NaI(Tl) detector is shown. This measurement was carried out over a period of 10 minutes. In the resulting spectrum, two specific peaks corresponding to \isotope[133]{Ba}, as detailed in Table~\ref{tab:Ba133}, are indicated with vertical dashed lines. Additionally, a peak related to \isotope[40]{K} from the natural background is marked in the same manner.}
	\label{fig:Ba133}
\end{figure*}

Using this method, we  estimated the activity of \isotope[133]{Ba}, as shown in Table~\ref{tab:Ba133}. These estimations are based on the measurement of two specific gamma rays (see Figure~\ref{fig:Ba133}). We can compare these estimated activity values with the \isotope[133]{Ba} activity obtained from the calibration certificate, which has been adjusted for time decay, resulting in a value of $3.92 \times 10^4$ Bq.

A study on the coincidence summing of X-rays and gamma rays from \isotope[133]{Ba} is discussed in a paper by \cite{novkovic2007coincidence}. The study focuses on tracking all decay paths and their results. Another paper by \cite{novkovic2009direct} demonstrated the use of sum--peak measurements for \isotope[133]{Ba} in direct activity measurements, achieving an activity accuracy better than 1\%. However, the detector system used in these studies has a much better energy resolution than the NaI(Tl) system. Additionally, the decay scheme of \isotope[133]{Ba} is more complex than that of \isotope[60]{Co}, leading to a more intricate system of count rate equations that was not addressed in the research. 

The activity values obtained from Table~\ref{tab:Ba133} are within 4\% (for the 303 keV gamma ray) and 11\% (for the 356 keV gamma ray) of the activity determined from the calibration certificate. These differences fall within the specified accuracy criterion for activity determination outlined in Section~\ref{sec:intr}.

The verification process with a known radioactive source can be carried out through direct measurements using the same geometry and nuclide (for example, \isotope[60]{Co}) as those used  for the multisource standard. It is important to utilize the same energy calibration, obtained from Equation~\ref{eq:en_ch_fit}, and the same time interval.

\begin{figure*}[t]
	\centering
	\includegraphics[width=\textwidth]{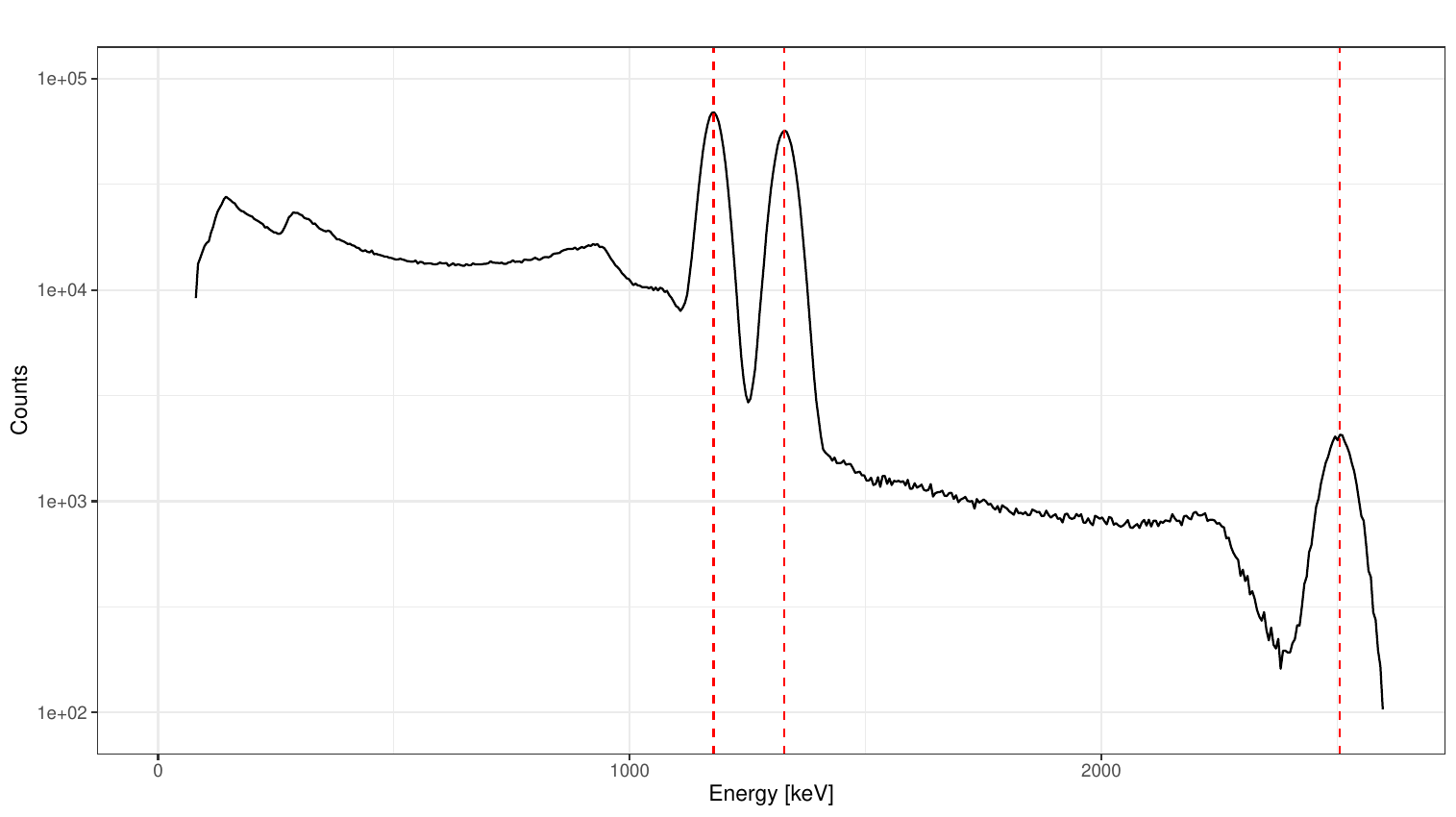} \\[-10pt]
	\caption{Spectrum of the \isotope[60]{Co} radionuclide standard measured via a NaI(Tl) detector. This measurement lasted for 10 minutes. In the spectrum obtained from this measurement, three distinct peaks associated with \isotope[60]{Co}, which are described in more detail in Table~\ref{tab:Co60}, are marked with vertical dashed lines.}
	\label{fig:Co60}
\end{figure*}

Figure~\ref{fig:Co60} displays the spectrum of the \isotope[60]{Co} radionuclide standard captured on May 23, 2023, with a total live time duration of 10 minutes.

\begin{table*}[t]
	\centering
	\caption{A summary of the ROI peaks observed in the spectrum of a \isotope[60]{Co} radionuclide standard is presented in Figure~\ref{fig:Co60}. The measurement was performed on May 23, 2023, starting at 10:46:34 AM, and lasted for a total live time of 10 minutes. }
    \begin{tabular}{cccccccccc}
    	\hline
    	\\ [-10pt]
    	Nuclide & ROI & ROI & $ N_m(E) $ & $ \sigma_i(E) $ & FWHM & $ E $ & Centroid & $ p_i $ & References \\
    	&     & Counts & Counts & Counts & keV  & keV & keV & $\times 100$ & \\
    	\hline
    	\\ [-10pt]
		\isotope[60]{Co} & 1 & 959,732 & 762,098 & 1,083 & 55.3 & 1,173 & 1,173 & 99.85 & \cite{TabRad_v3} \\ 
		\isotope[60]{Co} & 2 & 778,552 & 706,783 &   982 & 60.2 & 1,333 & 1,328 & 99.98 & \cite{TabRad_v3} \\ 
		\isotope[60]{Co} & 3 &  42,007 &  33,665 &   210 & 81.6 & 2,506 & 2,509 & $\sum$-peak &  \\
    	\hline
    \end{tabular}%
	\label{tab:Co60}%
\end{table*}%

Table~\ref{tab:Co60} provides a summary of the net area counts obtained from the \isotope[60]{Co} standard source in the same near-surface geometry and with the same live time as the multisource standard. By utilizing the information presented in the table and employing the ``\text{simplified}'' solution from Equation~\ref{eq:s_sol}, it is possible to estimate the activity of the \isotope[60]{Co} standard source. This estimation can then be compared with the activity of the standard source to assess the accuracy of the activity determination. The activity of the \isotope[60]{Co} standard source was determined to be $3.4(1) \times 10^4$~Bq. This value is within 9\% of the reported \isotope[60]{Co} standard activity stated in the calibration certificate~(\cite{unknown-author-2022}) and adjusted for the time of the measurements ($3.7(1) \times 10^4$~Bq). Therefore, the activity determination meets the required accuracy criterion of being within 15\%.

Importantly, the simplified solution can be used to determine the activity of \isotope[60]{Co} even if the activity is unknown. Alternatively, one can deduce the activity of the \isotope[60]{Co} standard source by measuring the net area counts and comparing them to the net area counts of the same photopeak from the calibration source, taking into account radioactive decay correction. On the basis of the information provided in Table~\ref{tab:mu_su} and Table~\ref{tab:Co60}, shows that  the estimated activity of the \isotope[60]{Co} standard source is $3.5(1) \times 10^4$~Bq ($ \frac{\text{762,098} \times 3.7}{\text{794,738}} \simeq 3.5 \times 10^4 $~Bq), which is within 5\% accuracy (approximately $ \frac{3.7 - 3.5}{3.7} \simeq 5\% $). In this case, the radioactive decay correction can be neglected since the multisource calibration measurements and \isotope[60]{Co} standard source measurements were conducted within one month of each other and the correction is negligible compared with the half-life of \isotope[60]{Co}.

\begin{figure*}[t]
	\centering
	\includegraphics[width=\textwidth]{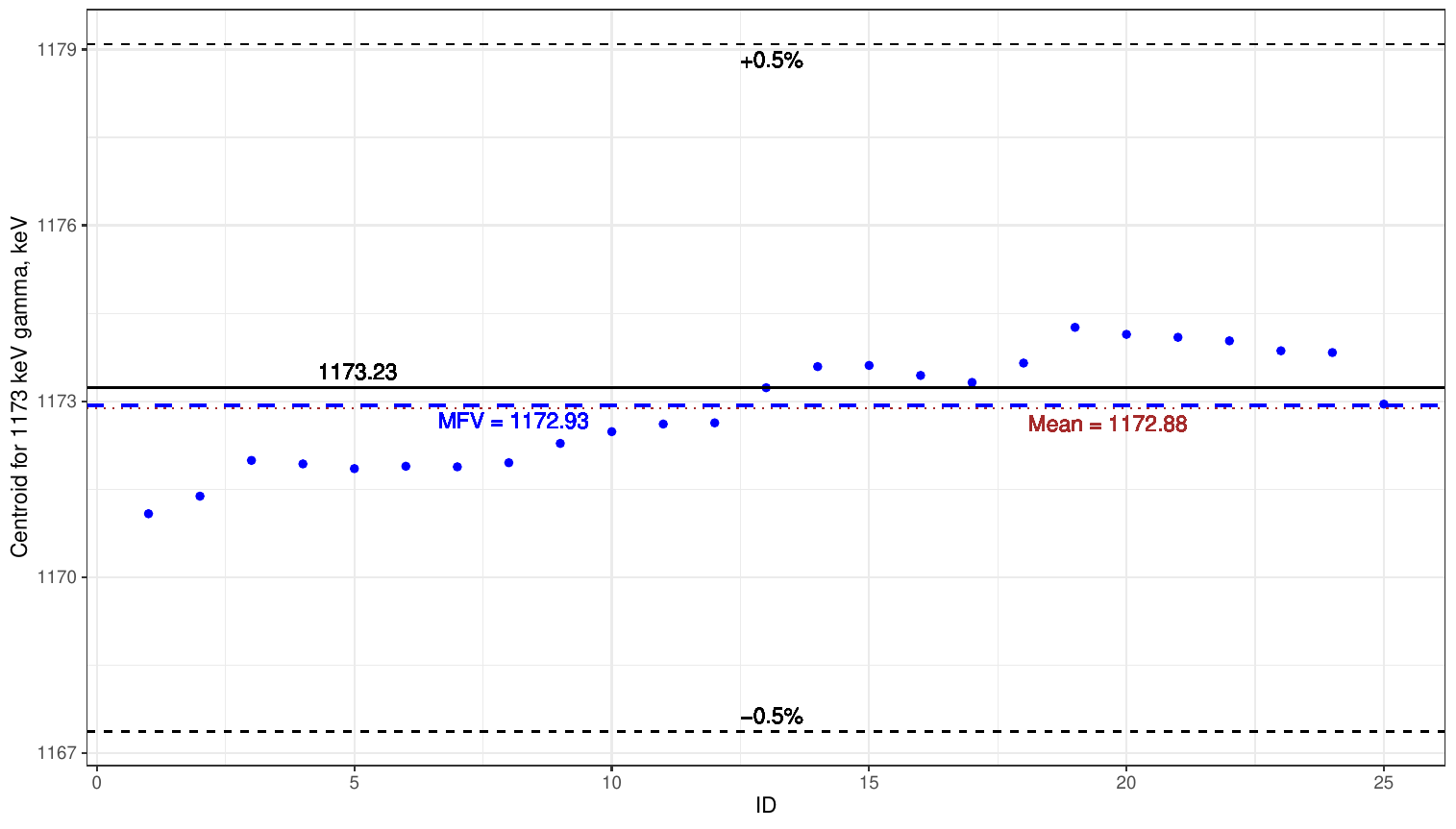}\\ [-10pt]
	\caption{The stability of the energy calibration can be seen in this graph. All 25 measured centroid values corresponding to \isotope[60]{Co} 1,173.23~keV gamma-rays clearly  fall within an uncertainty of  $\pm 0.5\%$.}
	\label{fig:Co60Centr}
\end{figure*}
The stability of the energy calibration can be seen in Figure~\ref{fig:Co60Centr}. This graph displays 25 measurements of the centroid values of 1,173.23~keV gamma rays from \isotope[60]{Co} over a few weeks, where the original energy calibration was obtained from a mix of radioactive nuclei (see Figure~\ref{fig:NaI_mult}). The black solid line marks the reference energy value at 1,173.23 keV. The two dashed lines represent 0.5\% deviations above and below this reference value. The graph includes a dashed line for the most frequent value (MFV) (see \ref{ap:MFV}) of the measurements, which is 1,172.93, and a dotted line for the average (mean) value, which is 1,172.88. The practical application of the MFV method and its advantages in analyzing different datasets were recently discussed in the references \cite{golovkoApplicationMostFrequent2023} and \cite{golovko2023unveiling}. 
The figure clearly shows  that the energy calibration stability of the sodium iodine detector is better than $\pm 0.5\%$, as observed from the centroid values for the 1,173.23~keV gamma ray.

\begin{figure*}[t]
	\centering
	\includegraphics[width=\textwidth,trim=0 0 0 16pt]{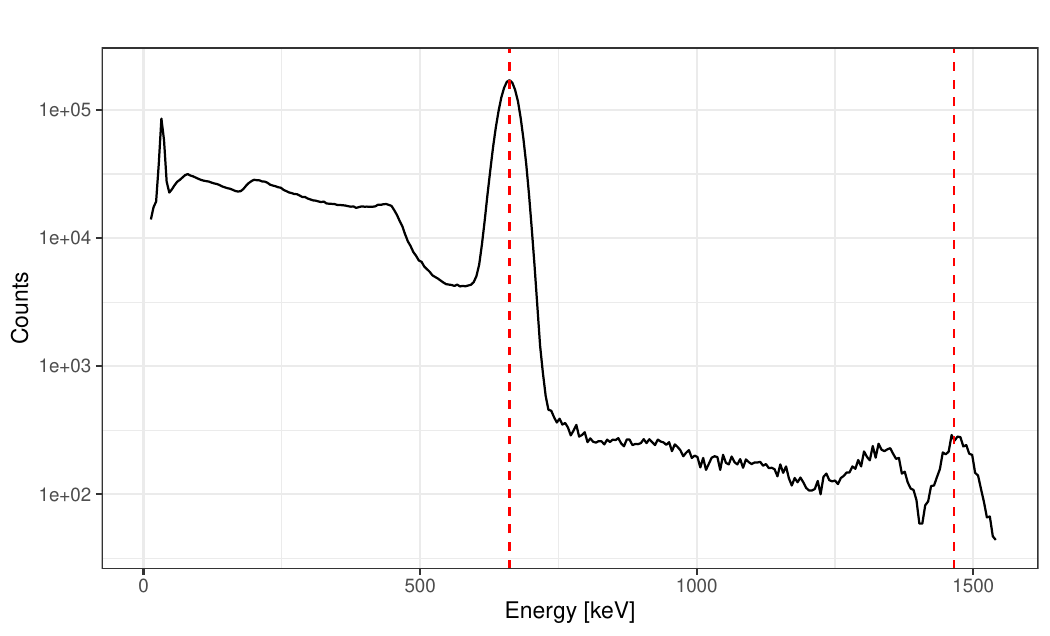} \\[-10pt]
	\caption{A 10-minute measurement of \isotope[137]{Cs} using a NaI(Tl) detector revealed  a 662 keV energy peak, marked with a dashed line, along with a peak for naturally occurring \isotope[40]{K}.}
	\label{fig:Cs137}
\end{figure*}
Figure~\ref{fig:Cs137} displays the spectrum of the \isotope[137]{Cs} radionuclide standard captured on May 30, 2023, with a total live time duration of 10 minutes.

\begin{figure*}[t]
	\centering
	\includegraphics[width=\textwidth]{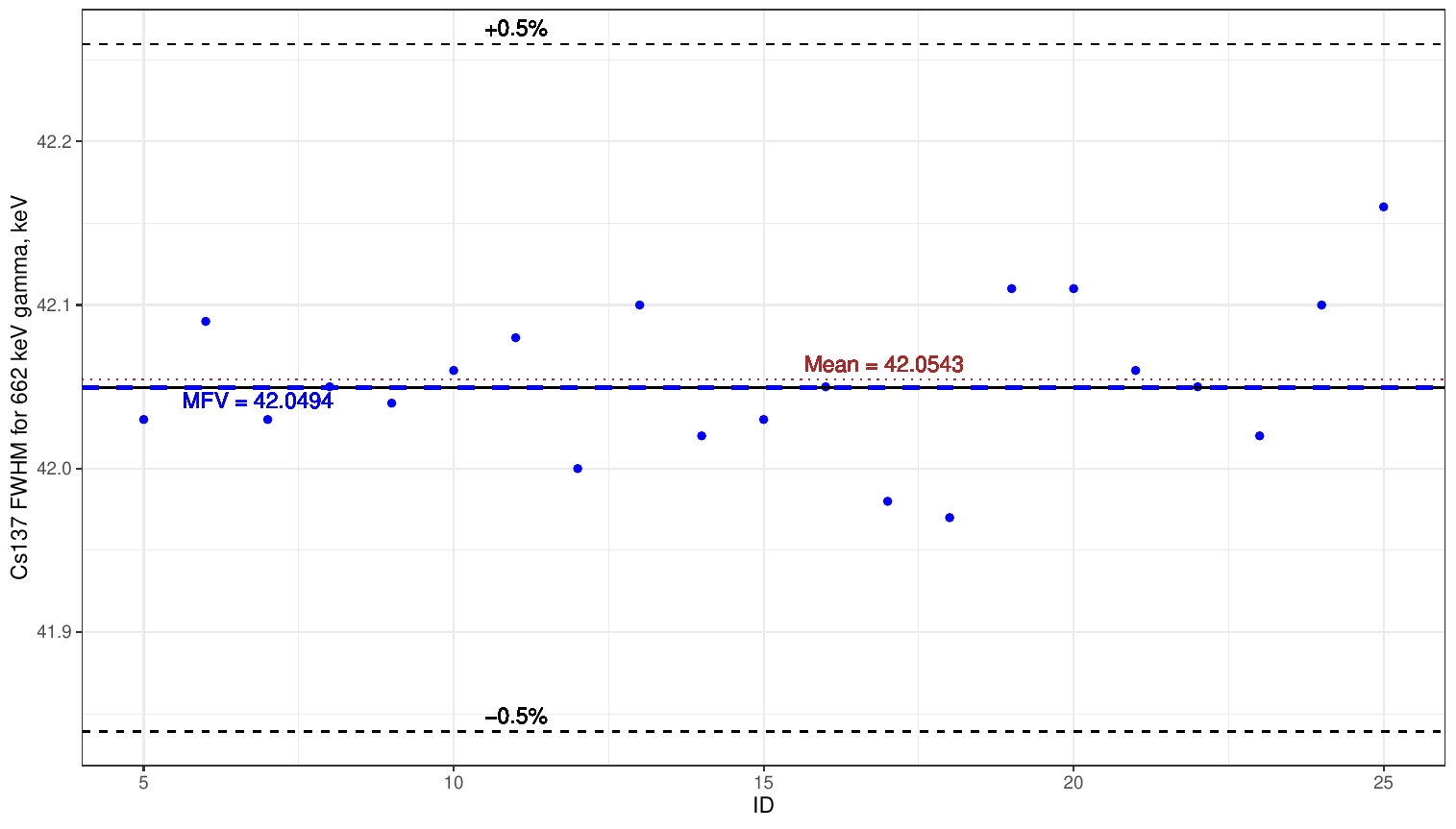} \\ [-10pt]
	\caption{The graph displays the stability of the FWHM for the 662~keV gamma rays of \isotope[137]{Cs}. These values are within an accuracy range of $\pm 0.5\%$. This indicates that the FWHM values remain consistent and stable, highlighting the reliable performance of the measurement system.}
	\label{fig:Cs137-FWHM}
\end{figure*}
Figure~\ref{fig:Cs137-FWHM} provides a graphical representation of the stability of the FWHM for the 662~keV gamma rays emitted by \isotope[137]{Cs}. The data clearly show  that all 21 measurements of the \isotope[137]{Cs} FWHM for the 662~keV gamma rays are within an uncertainty range of  $\pm 0.5\%$. This finding  indicates that the FWHM values remain stable and consistent for the given energy, demonstrating the reliability and precision of the NaI(Tl) detector. Figure~\ref{fig:Cs137-FWHM} shows  the representation of the mean and MFV of the FWHM dataset, as described in~\cite{golovkoApplicationMostFrequent2023}. Interestingly, the two values align with each other. Additionally, the figure shows  that the detector's resolution is better than 7\% (approximately 7\% considering $ \frac{42.05}{661.66} + 0.5\% $). This value is significantly lower than the acceptable level of 10\% for the NaI(Tl) detector's resolution for the characteristic 662~keV gamma ray emitted by \isotope[137]{Cs}.

\subsubsection{Detection Limits at Near-contact Geometry}
\label{sec:LLD}

Detection limits refer to the measurement system's ability to detect a minimum amount of activity for a particular gamma-emitting radionuclide under specific conditions. These limits can be estimated via various mathematical expressions. One widely accepted expression for estimating detection limits is known as the lower limit of detection (LLD). It incorporates a predetermined risk level of 5\% for falsely concluding the presence of activity and a 95\% level of confidence for detecting the actual presence of activity.

The calculation of detection limits is described in references such as \cite{pasternack1971detection} and \cite{head1972minimum}. These references provide methods and formulas to determine the lowest detectable activity for a given radionuclide at the time of measurement. The LLD is a way to check how sensitive and capable a measurement system is at spotting small amounts of radioactivity. This ensures that there is a 95\% chance of accurately detecting whether radioactivity is present. The equation is as follows:
\begin{equation}
	\text{LLD} = \frac{4.66 \cdot \sigma_i}{ \varepsilon_i(E) \cdot p_i }  ,
	\label{eq:LLD}
\end{equation}
where $\sigma_i $ is the estimated standard error of the net count rate, $\varepsilon_i(E) $ is the counting efficiency of the specific nuclide's energy (value $ \le $1), $p_i $ is the absolute transition probability by gamma decay through the selected energy (value  $ \le $1), and $ E $ is the energy (keV) of the corresponding gamma line. 

The LLD described in Equation~\ref{eq:LLD} allows us to assess the performance of a gamma-measuring system independently from the sample being measured. It assumes that the count rate in the energy range corresponding to the specific nuclide and the count rate in the background region(s) are roughly equal. This equation provides a way to estimate the system's ability  to detect low levels of radioactivity by considering the count rates in specific energy areas and background regions.

When a detector records activity levels close to or below its LLD, it indicates very low radioactivity, often near the system's minimum capability to detect. It is essential to check the accuracy, assess the relevance of such low levels, consider using more sensitive equipment if necessary, and interpret results cautiously due to the limitations at these low detection thresholds.

\begin{table*}[htbp]
	\centering
	\caption{Estimated LLD for each single radionuclide used in this study, except for \isotope[241]{Am}, which was obtained from a mixed radionuclide source. The estimated activity is calculated via Equation~\ref{eq:exp_coef}. The table includes the nuclide activity from the calibration certificates, allowing for a comparison of the percent difference (Diff.) between the activity estimation ($S$ Est.) and the certified value ($S$ Cert.). The net counts for a specific gamma ray were measured over a duration of 10 minutes of live time.}
    \begin{tabular}{rrccccccrc}
    	\hline
    	\\ [-10pt]
    	Nuclide & $ E $ & $ N_m(E) $ & $ \sigma_i(E) $ & $ \varepsilon_i(E) $ & $ p_i  $  & $ S $ Est. & $ S $ Cert. & Diff. & LLD \\
    	& keV & Counts & Counts &  & $ \times 100 $  & Bq & Bq &  & Bq \\
    	\hline
    	\\ [-10pt]
		$^{241}$Am &    59.54 & 1,305,782 & 1,355 & 0.1388 & 35.92 & 43,654 & 43,658 & 0.01\% & 211 \\ 
		$^{57}$Co &   122.06 & 2,345,499 & 1,765 & 0.1870 & 85.60 & 24,418 & 24,437 & 0.08\% &  85 \\ 
		$^{133}$Ba &   302.85 &   560,175 & 1,174 & 0.1376 & 18.31 & 37,052 & 39,196 & 5.47\% & 361 \\ 
		$^{133}$Ba &   356.02 & 1,578,425 & 1,461 & 0.1225 & 62.05 & 34,596 & 39,196 & 11.74\% & 149 \\ 
		$^{137}$Cs &   661.66 & 1,537,147 & 1,525 & 0.0696 & 85.10 & 43,278 & 43,105 & 0.40\% & 200 \\ 
		$^{60}$Co & 1,173.23 &   762,098 & 1,083 & 0.0367 & 99.85 & 34,651 & 37,639 & 7.94\% & 229 \\ 
		$^{60}$Co & 1,332.49 &   706,783 &   982 & 0.0316 & 99.98 & 37,289 & 37,639 & 0.93\% & 241 \\
    	\hline
    \end{tabular}%
	\label{tab:LLD}%
\end{table*}%

Table~\ref{tab:LLD} presents the LLDs for the nuclides used in this study, specifically in the near-contact geometry. The measurement was conducted for a duration of 10 minutes. The calculation of the LLD was performed via Equation~\ref{eq:LLD}. The certified activity values listed in the table were obtained from~\cite{unknown-author-2022} and adjusted for radioactive decay. The estimated activity was determined via Equation~\ref{eq:activity}, and the efficiency fit results from Equation~\ref{eq:exp_coef}. The table includes the difference between the estimated activity and  certified activity, demonstrating that it falls within the acceptance criterion of 15\%.

\section{Discussion}

The ``\text{oversimplified}'' and ``\text{simplified}'' models show a good match between gamma spectrometry readings and radiation levels from a calibrated \isotope[60]{Co} source. The ability to  calculate the efficiency calibration for a single-crystal scintillation detector system easily can be very useful in nuclear cleanup efforts, especially if a weak \isotope[60]{Co} source is discovered in the field. By applying true coincidence summing methods, one can then estimate the efficiency of the 1,173 and 1,333 keV gamma-ray peaks in the decay of \isotope[60]{Co} with an accuracy that is better than 15\%. This approach  provides a reliable estimation of the efficiency of these gamma-ray peaks.

The best way to address the issue of true coincidence summing corrections is to calibrate the detector using a standard source of the nuclide being studied. By doing so, we can avoid having to account for coincidence summing effects. When calculating the photopeak efficiency of two gamma-ray lines in the decay of \isotope[60]{Co}, it is possible to use uncomplicated true coincidence summing correction methods without needing to know the source activity precisely. In our study, we utilized this approach to assess the accuracy of the simplification in the true coincidence summing method itself. Importantly,  angular correlation affects  the decay of \isotope[60]{Co} when  the true coincidence summing method is applied to achieve an accuracy of~15\%. Notably, when  the ``\text{simplified}'' model is used for estimating the activity of \isotope[60]{Co} (refer to Equations~\ref{eq:simSolDatEr} and \ref{eq:dec_data_er}), the difference in the estimated activity is greater than when  the ``\text{general}'' solution is used (refer to Table~\ref{tab:LLD}). 

In addition, we  confirmed that the ``\text{general}'' solution for near-contact geometry can achieve an accuracy of 15\% in estimating activity via a single-crystal NaI(Tl) scintillator detector. This level of accuracy is considered sufficient for conducting field screening during nuclear remediation activities. 

\hl{In numerical analysis, there are two main types of algorithms used for solving systems of equations: explicit (see}~\ref{ap:theory}) \hl{and implicit (see}~\ref{ap:MFV}). \hl{Explicit algorithms calculate the solution directly  on the basis of the  current state of the system, whereas  implicit algorithms solve for the future state by considering both the current and future states simultaneously. In this study, we used the following explicit solution methods: ``oversimplified,'' ``simplified,'' and ``general.'' The computational cost and resource consumption for these methods are low. For example, the computational time is less than 1 second, and the memory consumption is less than 1 megabyte for the ``simplified'' algorithm running on a Windows PC with a 13th generation Intel(R) Core(TM) i9-13950HX processor at 2.20~GHz.}

\hl{The universality and adaptability of the algorithms are demonstrated through their application in the calibration of NaI(Tl) scintillation detectors in this study,  which is aimed at cleaner production and environmental protection. Moreover, a previous study showed that these algorithms have been adopted for the calibration of semiconductor detectors used in criticality dosimetry}~(\cite{golovko2022simplified}). \hl{The methods were also applied by}~\cite{massarczyk2023constraints} \hl{to normalize the absolute efficiency determined from the Monte Carlo-derived efficiencies of semiconductor detectors. While we have adapted the algorithms for use with scintillation detectors, further exploration of their adaptability across a broader range of applications in physics could be the focus of future work.}

\hl{Section}~\ref{sec:Ener_Cal} \hl{discusses the ``oversimplified'' and ``simplified'' techniques and their accuracy for NaI(Tl) detectors, comparing them to the ``general'' technique, which is commonly used in the remediation field. The adaptation of these techniques for scintillation detectors significantly reduces the operational costs of nuclear remediation work. Typically, semiconductor detectors with the same sensitivity as NaI(Tl) are at least ten times more expensive. Additionally, the operation of high-purity germanium detectors requires cooling the germanium crystal to liquid nitrogen temperatures, which takes much longer than  NaI(Tl) crystals,  which   can operate at room temperature. Furthermore, measuring the true coincidence summing photopeak in the decay of $^{60}$Co allows  purchase of a calibrated $^{60}$Co source, as required in the ``general'' technique, to be avoided. A subpercent calibrated $^{60}$Co standard source is very expensive.}

\hl{Section}~\ref{sec:Ener_Cal} \hl{validates the long-term operation  stability  and reliability of the NaI(Tl) detector system used for nuclear remediation. Using the 1,173.23~keV gamma-rays from a calibrated $^{60}$Co source, we have demonstrated that the measured centroid values remain within a $\pm$0.5\% uncertainty over the long term. Additionally, we have shown that the resolution of the NaI(Tl) detector system is stable over time by measuring the full width at half maximum  for the 662~keV gamma-rays of $^{137}$Cs.}

The single-crystal sodium iodine detector met all the acceptance criteria specified in Section~\ref{sec:intr} when placed in a near-contact geometry configuration. Additionally, we conducted tests on the system's performance using a 15 cm geometry configuration that was not discussed in this paper, and it also passed all acceptance criteria for that setup. These results indicate that the NaI(Tl) detector is effective and appropriate for applications in nuclear remediation. The relevant data are available in the online repository \href{https://osf.io/z352h}{https://osf.io/z352h} (\cite{osf2023dataset}).

\section{Conclusion}

In the near-contact geometry configuration, the device under test  successfully met all the specified acceptance criteria outlined in  Section~\ref{sec:intr}. Additionally, a low limit of detection was determined for all the nuclides used in the calibration process for this specific geometry. It has been demonstrated that the ``\text{simplified}'' and ``\text{oversimplified}'' approximations for estimating the activity of the \isotope[60]{Co} nuclide are applicable to the single-crystal sodium iodine detector (in this work) as well as for high-purity germanium detectors~(\cite{golovko2022simplified}). These approximation methods (as well as the ``\text{general}'' solutions) yield an accuracy that falls within the acceptance criterion for the NaI(Tl) detector, making them suitable for use as  screening tools for nuclear remediation purposes. The stability of the energy calibration and the energy resolution of the NaI(Tl) detector have also been verified. 

Furthermore, we have demonstrated that the innovative methods previously employed to commission high-purity germanium detectors~(\cite{golovko2022simplified}) can be adapted for commissioning scintillation detectors used in remediation. Although NaI(Tl) detectors typically have poorer energy resolution than semiconductor technology does, they can still be calibrated via a ``simplified'' and ``oversimplified'' approximations for estimating the activity of the $^{60}$Co nuclide.

This work is  aimed mainly at showing that  {\rm ``oversimplified''} and {\rm ``simplified''} methods for coincidence summing corrections in NaI(Tl) spectrometry result in an accuracy within acceptance criteria, which is sufficient in most fields of application for nuclear remediation screening. The NaI(Tl) detector system at the CRL can be used for nuclear remediation. Moreover, suggested and verified methods based on the true coincidence summing effect, for uncomplicated determination of the photopeak efficiency of the scintillator NaI(Tl) detector system, are applicable for use with any NaI(Tl) detector system. This method and calibrated  radioactive sources can be used to commission any other NaI(Tl) detector for use during nuclear remediation. 

Additionally, a noncommercial data analysis tool such as R was utilized in this work. This programming language is open-source and free, making it possible to read raw data files from Canberra detectors and analyze the data. Using R can lead to cost savings in nuclear remediation activities by avoiding the high license fees often associated with commercial software. Furthermore, R is embraced by a wide scientific community and subjected to rigorous verification and validation processes beyond those typically seen in the smaller nuclear physics community. This ensures its reliability and accuracy for data analysis.

In summary, the ``oversimplified'' method and ``simplified'' method are well suited for the efficient calibration of the NaI(Tl) detector system for use during nuclear remediation. The inclusion of  angular dependence in  detector efficiency estimation is an important part of simplified methods. Simplification in determining detector efficiency and source activity, with accuracy within a few percent, shows that although simple, these methods are well suited for efficiency calibration of scintillation detectors. The research verified this through direct comparison of activity determination from a calibrated source. The validation using calibrated radioactive sources demonstrates that the methods have been thoroughly tested and verified. The stability of the energy calibration and the energy resolution of the NaI(Tl) detector were also verified. The successful commission of the NaI(Tl) scintillation detector system at the CRL suggests its suitability for screening radioactive isotopes at contaminated sites, highlighting the practical application and reliability of these novel methods.

\section*{Acknowledgment}

I would like to extend my gratitude to Genevieve Hamilton and David Yuke for their invaluable support of this work. Their encouragement has been instrumental in the successful completion of this project. Additionally, I must acknowledge the contributions of Alan MacDonald, Jason Thibeault, Thomas Wilson, Ayron O'Grady, Kevin Campbell, Greg Hersak, Kurt Jensen, and Anthony Rossi. Each has provided vital support and insights that have greatly enriched this work. Their dedication and assistance have been truly appreciated and have played a significant role in achieving our objectives.
I am grateful to Helena Rummens for her careful final editing of the paper.

\section*{Data Availability Statement} 
This manuscript has associated data available in the following repository: \href{https://osf.io/pbmd2/}{https://osf.io/pbmd2/} (\cite{osf2023dataset}).

\section*{Declaration of generative AI and AI-assisted technologies in the writing process}
During the preparation of this work the author used ChatGPT  to check the language of the manuscript. After using this tool, the author reviewed and edited the content as needed and takes full responsibility for the content of the publication.

\newpage
\section*{Abbreviations}{
	The following abbreviations are used in this manuscript:\\
	
	\noindent 
\begin{tabular}{@{}ll} %\begin{tabular}{p{3.5cm}p{12 cm}}
	Ch  & channel number \\
	cps & counts per seconds \\
	CRL & Chalk River Laboratories\\
	FWHM & full width at half maximum \\
	LLD & lower limit of detection \\
	MFV & most frequent value \\
	NaI(Tl) & thallium-activated sodium iodide \\
	ROI & region of interest \\
\end{tabular}
}

\appendix
\section{Theory of the sum--peak method}
\label{ap:theory}

We  discuss a situation where two photons occur  at the same time, such as in the decay of \isotope[60]{Co}. We do not consider adjustments for pileup pulses because we believe that the radioactive sources of  legacy nuclear waste sites  are weak. This is because the level of radioactive contamination is predicted to be low during the nuclear cleanup process.  

The full-energy peaks areas $N_1$,  $N_2$, and $N_{12}$, under the photopeaks and sum--peak, respectively, of an isotope that emits  two gamma rays in coincidence are given by \cite{brinkman1963absolute}, \cite{MALONDA2012871}, \cite{MALONDA2020531}, and \cite{golovko2022simplified}:
\begin{equation}
	\label{eq:sp_th}
	\begin{split}
		N_1 & = \varepsilon_1 p_1 A \left( 1- \eta_2 p_2 \omega(0^{\circ}) \right) \\ 
		N_2 & = \varepsilon_2 p_2 A \left(1- \eta_1 p_1 \omega(0^{\circ}) \right) \\
		N_{12} & = \varepsilon_1 p_1 \varepsilon_2 p_2 A \omega(0^{\circ}), 
	\end{split}
\end{equation}
where $A$ is the absolute activity of the source, $\varepsilon_1$, and $\varepsilon_1$ are the efficiencies for the full-energy peak,  $\eta_1$ and $\eta_1$ are the total efficiencies, and $p_1$ and $p_2$ are the photon emission intensities of gamma rays per disintegration. $\omega(0)$ is the correction factor for angular correlation. 

In the {\rm ``oversimplified''} method solution described in Equation~\ref{eq:ovs_sol}, it is assumed that the total efficiencies in Equation~\ref{eq:sp_th} are not considered. This means that the total efficiencies mentioned in Equation~\ref{eq:sp_th} are not considered in this particular case and that Equation~\ref{eq:sp_th} will be transformed as follows:  
\begin{equation}
	\label{eq:sp_thos}
	\begin{split}
		N_1 & = \varepsilon_1 p_1 A  \\
		N_2 & = \varepsilon_2 p_2 A  \\
		N_{12} & = \varepsilon_1 p_1 \varepsilon_2 p_2 A \omega(0^{\circ}). 
	\end{split}
\end{equation}
This simplification helps  make the calculations easier and quicker.

In the {\rm ``simplified''} method solution described in Equation~\ref{eq:s_sol}, we assume that the total efficiencies in Equation~\ref{eq:sp_th} can be replaced by the efficiencies for the full-energy peak, which means that $\eta = \varepsilon$, and Equation~\ref{eq:sp_th} can be transformed as follows: 
\begin{equation}
	\label{eq:sp_ths}
	\begin{split}
		N_1 & = \varepsilon_1 p_1 A \left(1- \varepsilon_2 p_2 \omega(0^{\circ}) \right)  \\
		N_2 & = \varepsilon_2 p_2 A \left(1- \varepsilon_1 p_1 \omega(0^{\circ}) \right) \\
		N_{12} & = \varepsilon_1 p_1 \varepsilon_2 p_2 A \omega(0^{\circ}). 
	\end{split}
\end{equation}
This simplification implies that only photoelectric absorption processes are taken into account for this specific case. Equation~\ref{eq:sp_ths} can be expressed differently, as shown  in Equation~\ref{eq:sp_thss}:
\begin{equation}
	\label{eq:sp_thss}
	\begin{split}
		N_1 + N_{12} = \tilde{N_1} & = \varepsilon_1 p_1 A \\
		N_2 + N_{12} = \tilde{N_2} & = \varepsilon_2 p_2 A \\
		N_{12} & = \varepsilon_1 p_1 \varepsilon_2 p_2 A \omega(0^{\circ}). 
	\end{split}
\end{equation}
This new form is quite similar to the {\rm ``oversimplified''} Equation~\ref{eq:sp_thos}. Consequently, the solution for the {\rm ``simplified''} case closely resembles the solution for the {\rm ``oversimplified''} case, as demonstrated in Section~\ref{sec:ovs} and Section~\ref{sec:s}.

\section{The most frequent value of the dataset and its variance}
\label{ap:MFV}

This is a brief overview of a robust method for determining the central value in a dataset, which was created by Steiner and outlined in detail in \cite{Steiner1973}, \cite{Csernyak1973}, \cite{Ferenczy1988ShortIntroduction}, \cite{steinerMostFrequentValue1988}, \cite{steinerIntroductoryInstructionsComputations1991}, \cite{steinerMostFrequentValue1991}, \cite{steinerOptimumMethodsStatistics1997}, and \cite{kemp2004steiner}.

To find the most common value in a dataset ($x_1,...,x_i,...,x_N$), we can calculate the MFV ($M$) and the scale parameter (denoted by $\epsilon$ and called dihesion) iteratively. This task involves calculating a specific mathematical expression multiple times. The expression is as follows:
\begin{equation}
	M_{j+1} = \frac{ \sum_{i=1}^{N} x_i \cdot \frac{\epsilon^2_j}{\epsilon^2_j + \left( x_i - M_j \right)^2 } }{ \sum_{i=1}^{N} \frac{\epsilon^2_j}{\epsilon^2_j + \left( x_i - M_j \right)^2} }.
	\label{Eq:MFV}
\end{equation}
In this equation, $\epsilon_j$ is also calculated as a result of an iterative process that is performed a certain number of times. The formula for this calculation is  as follows:
\begin{equation}
	\epsilon^2_{j+1} = 3 \cdot \frac{ \sum_{i=1}^{N} \frac{( x_i - M_j)^2 }{ \left( \epsilon^2_j + \left( x_i - M_j \right)^2 \right)^2 } }{ \sum_{i=1}^{N} \frac{ 1 }{ \left( \epsilon^2_j + \left( x_i - M_j \right)^2 \right)^2 } }.
	\label{Eq:dihesion}
\end{equation}

To start the iterations ($j=0$), we first set the initial values. We set $M_{(0)}$ as the mean of the dataset, which is calculated as $\frac{1}{N} \sum_{i=1}^{N} x_i$. 
It is recommended to start with the median $M_{(0)}$ instead of the mean when performing the MFV iterations (\cite{HajagosSteiner1992}). This is because the mean can be heavily influenced by outliers or the tails of the distribution, which could require many additional iteration steps to reach the desired result. In cases where the initial value of $\epsilon_j$ is significantly larger than the correct value, the number of iteration steps on the $\epsilon_j$-branch may need to be increased significantly.

The study by \cite{Hajagos1980} discusses a method to find the value of $\epsilon_{(0)}$ efficiently for rapid calculations of the MFV. In simple terms,  we can set $\epsilon_{(0)} = \frac{\sqrt{3}}{2} \cdot (x_{\text{max}} - x_{\text{min}})$, where $x_{\text{max}}$ and $x_{\text{min}}$ are the highest and lowest values in the dataset used to estimate the MFV. This formula helps in determining $\epsilon_{(0)}$ accurately for the MFV calculations.
In the process, we need to ensure that both Equations~\ref{Eq:MFV} and \ref{Eq:dihesion}, which are used to compute the values of $M_{j+1}$ and $\epsilon^2_{j+1}$ from the datasets, are satisfied simultaneously.

The calculation of  the empirical value of $M_{j+1}$ can be computationally intensive. Various practical applications of this method have been discussed in the works of \cite{FerenczySteiner1988MostFrequentValues}, \cite{szucsApplicabilityMostFrequent2006}, \cite{zhangMostFrequentValue2017}, \cite{szaboMostFrequentValuebased2018}, \cite{zhang2018most}, \cite{zhang2022mfv}, \cite{golovkoApplicationMostFrequent2023},  \cite{golovko2023unveiling}, and \cite{zhang2024most}. 

\cite{CsernyakSteiner1983} provided a simple formula for calculating the variance $\sigma_{M_j}$ at each iteration step in a symmetrical distribution when $M_{j+1}$ is determined according to Equation~\ref{Eq:MFV}:
\begin{equation}
	\sigma_{M_j} = \frac{\epsilon_j}{\sqrt{n_{\text{eff}}}}.
	\label{Eq:sig_M}
\end{equation}
In this formula, $\varepsilon$ represents the dihesion (which is the convergence value of the iterations as defined in Equation~\ref{Eq:dihesion}), and $n_{\text{eff}}$ is the effective number of data points. The number of effective points influencing the MFV result (\cite{Csernyak1980, HajagosSteiner1992}) is calculated as:
\begin{equation}
	n_{\text{eff}} = \sum_{i=1}^{N} \frac{\epsilon_j^2}{\epsilon_j^2 + \left( x_i - M_j \right)^2}.
\end{equation}
This formula offers a way to determine the variance that characterizes the accuracy when  $M$ is estimated for symmetrical distributions. 

The R and ROOT (\cite{brun1997root}) codes calculating the most frequent value and its variance for the symmetrical data is located in the following repository \href{https://osf.io/pbmd2/}{https://osf.io/pbmd2/} (\cite{osf2023dataset}).

\end{document}